\documentclass[11pt]{article}
\usepackage{color}
\usepackage{amsmath,amsfonts,amssymb,stackengine,graphicx,epsfig,latexsym}
\usepackage{fullpage}
\usepackage[T1]{fontenc}
\usepackage{hyperref}
\usepackage{amssymb,amsmath}

\newcommand{\be}{\begin{eqnarray}}
\newcommand{\ee}{\end{eqnarray}}

\newcommand{\planck}{P{\scshape lanck }}

\usepackage{cite}

\begin{document}

\setcounter{footnote}{0}

\baselineskip 6 mm

\begin{titlepage}
	\thispagestyle{empty}

{\ } 
	
	\vspace{35pt}
	
	\begin{center}
	    { \Large{\sc Non-Linear Supergravity and Inflationary Cosmology  }}
		
		\vspace{50pt}
		
		{Ignatios Antoniadis$^{1,2}$,  Emilian~Dudas$^{3}$, Fotis~Farakos$^{4}$ and Augusto~Sagnotti$^{5}$}
		
		\vspace{25pt}

		  {\it $^{1}$ High Energy Physics Research Unit, Faculty of Science, \\ Chulalongkorn University, Bangkok 1030,
					Thailand}
		
		\vspace{15pt}

        {\it $^{2}$ Laboratoire de Physique Th\'eorique et Hautes Energies - LPTHE \\
			Sorbonne Universit\'{e}, CNRS, 4 Place Jussieu, 75005 Paris, France}
		
		\vspace{15pt}

        {\it $^{3}$ CPHT Ecole Polytechnique, CNRS, IP Paris, 91128 Palaiseau, France}
		
		\vspace{15pt}

        {\it $^{4}$ Physics Division, National Technical University of Athens \\
        15780 Zografou Campus, Athens, Greece}
		
		\vspace{15pt}

        {\it $^{5}$ Scuola Normale Superiore and INFN \\ Piazza dei Cavalieri 7, 56126 Pisa, ITALY}
		
		\vspace{15pt}

		\vspace{40pt}
		
		{ABSTRACT}
	\end{center}

We review a series of developments concerning non-linear realizations of supersymmetry coupled to supergravity, with emphasis on some applications to inflationary Cosmology.

\vspace{10pt}

\begin{center}
\emph{Invited contribution to ``Half a century of Supergravity'' \\ eds.~A. Ceresole and G.~Dall'Agata (Cambridge Univ. Press, to appear)}
\end{center}

\bigskip

\end{titlepage}

\newpage

\tableofcontents

\newpage

\section{\sc Introduction}

When a vacuum breaks some symmetries of a system, non-linear realizations can efficiently capture the behavior of small fluctuations around it.
This principle surfaces, here and there, in Field Theory and Particle Physics, and supersymmetry is no exception~\cite{Volkov:1973ix,Rocek:1978nb,IK,Lind,Kapustnikov:1981de,Samuel:1982uh,Kuzenko:2011tj,Antoniadis:2010hs}: even in this context,
non-linear realizations can provide effective low-energy descriptions after the decoupling of heavy states ~\cite{Casalbuoni:1988xh,Komargodski:2009rz,Kuzenko:2010ef,Antoniadis:2011xi,Farakos:2013ih,Ferrara:2016een,DallAgata:2016syy,Cribiori:2017ngp}.

However, there is an intriguing setting in String Theory \cite{stringtheory}, usually dubbed ``brane supersymmetry breaking'' (BSB),  where non--linear realizations of supersymmetry seem to play a more fundamental role~\cite{sugimoto,Antoniadis:1999xk,Angelantonj:1999jh,Angelantonj:1999ms,Dudas:2000nv,Pradisi:2001yv}. The need for non--linear realizations is clearly revealed, in these cases, by the string spectra, whose rigidity forbids a corresponding linear regime, but the low--energy couplings can be nicely addressed nonetheless in supergravity~\cite{sugra,sugra_rev}, following~\cite{Dudas:2000nv,Pradisi:2001yv}.
``Brane supersymmetry breaking'' emerged in a class of orientifold constructions~\cite{orientifolds1,orientifolds2,orientifolds3,orientifolds4,orientifolds5,orientifolds6,orientifolds7,orientifolds8,orientifolds_rev} where the vacuum hosts branes and orientifolds that break complementary portions of the original supersymmetry. No tachyons are thus generated, in sharp contrast to what would happen with collections of nearby branes and antibranes, precisely because the orientifolds are not dynamical objects. Still, these extended objects exert a sizable back-reaction, so that in the simplest ten--dimensional setting of~\cite{sugimoto} a string--frame runaway potential of the form
\be
V \ = \ T \ \int d^{10}x \ \sqrt{-g} \ e^{-\,\phi}  \label{tadpole}
\ee
emerges from (projective) disk amplitudes. This implies that the original flat space is not a vacuum and, as we shall see, the precise value of the string--frame exponent, which is fixed by the Polyakov expansion, has striking consequences. Non-linear supersymmetry also plays an important role in the world-volume of D-branes that break half of the bulk linear supersymmetry, while restricting their low--energy effective field theory~\cite{Antoniadis:2001pt,Antoniadis:2004uk, Antoniadis:2008uk,Ambrosetti:2009za}.

Cosmological backgrounds break supersymmetry and afford embeddings in four--dimensional N=1 supergravity~\cite{sugra_rev},
a low-energy limit of String Theory. In the settings that we highlighted they are naturally addressed
relying on non-linear realizations. This review is devoted to illustrating some progress that has been made in these directions during the last few years.

The use of non-linear supersymmetry for inflationary model building was initiated in~\cite{Antoniadis:2014oya}, where the constrained superfield formalism led to a non--linear supergravity description of Starobinsky's model of inflation \cite{Cecotti:1987sa,Kallosh:2013lkr,Farakos:2013cqa}. 
The elimination of a heavy complex scalar (the sgoldstino) from the spectrum thus resulted in an effective low--energy description independent of ultraviolet details related to it. 
This work was followed by further developments~\cite{DallAgata:2014qsj,Kallosh:2014hxa,Scalisi:2015qga,Ferrara:2014kva,Ferrara:2015gta,Kahn:2015mla,Ferrara:2015tyn,Carrasco:2015iij,DallAgata:2015zxp,McDonough:2016der,Kallosh:2019apq} that explored the versatility of the non--linear setup in connection with model building, together with other articles aiming at more detailed cosmological setups~\cite{Hasegawa:2017hgd,Dalianis:2017okk,Dudas:2021njv,Bonnefoy:2022rcw}.
In the bulk of this review we shall focus on the properties of these classes of supergravity models, referring to some of the most relevant examples.

Non-linear supersymmetry can also be relevant for late-time Cosmology, but due to space limitations we can barely mention some developments along these lines.
A de Sitter vacuum provides a simple model for dark energy, and non-linear supersymmetry is a clean way to obtain it within supergravity, as discussed in~\cite{Dudas:2015eha,Bergshoeff:2015tra,Hasegawa:2015bza,Kuzenko:2015yxa,Antoniadis:2015ala,Hasegawa:2015era,Bandos:2015xnf}.
The string origin of four--dimensional non-linear supersymmetry has been investigated in a series of related works~\cite{Bergshoeff:2015jxa,Dasgupta:2016prs,Vercnocke:2016fbt,Kallosh:2016aep,Bandos:2016xyu,Aalsma:2017ulu,Kallosh:2017wnt,GarciadelMoral:2017vnz,Cribiori:2019hod,Parameswaran:2020ukp},
but  progress is hampered, in this case, by the lack of consensus on the actual nature of dark energy, both observationally and microscopically, as reviewed in~\cite{Danielsson:2018ztv}.

\section{\sc Cosmology after \planck and supergravity}

In this section we review some developments in observational cosmology and supergravity model building that brought non-linear supersymmetry closer to inflation~\footnote{Here and in the following sections we adopt the conventions of~\cite{Wess:1992cp}.}.

\subsection{\sc \planck constraints on inflation}

Inflation was originally conceived to account for the isotropy and homogeneity of our Universe~\cite{Lyth:1998xn,Mukhanov:2005sc}. If this primordial phase dominated by an effective cosmological constant had occurred, an exponentially fast acceleration would have
diluted various types of unwanted/unobserved relics (magnetic monopoles or other defects), along with inhomogeneities in matter and radiation densities, and a final reheating phase would have populated the Universe with matter. Most importantly, the quantum fluctuations of an inflationary background would have provided the seeds for structure formation~\cite{cm}. 

It was realized early on that inflation would have left peculiar imprints on the Cosmic Microwave Background (CMB), and most notably a tilt in its primordial power spectrum of scalar perturbations proportional to ${k}^{n_s-1}$, which we now observe. It is not the only possible mechanism, but the detection of primordial gravitational waves~\cite{Starobinsky:1980te}, which for the moment remain elusive, would make its case much stronger. The \planck collaboration has provided a precise determination of $n_s$ \cite{Planck:2018jri}, 
\be
n_s \ = \ 0.9665 \ \pm \ 0.0038 \ 
\mathrm{at \  68\% \ CL} \ ,
\ee
but so far there is only an upper bound on the ratio $r$ between the power spectra of tensor and scalar perturbations,
\be
 r \ < \ 0.032 \ \mathrm{at \  95\% \ CL} \ .
\ee

The simplest model that can yield an inflating background involves a scalar field $\phi$, called the ``inflaton'', minimally coupled to gravitation, with a Lagrangian of the form
\be
\label{GR+INF}
e^{-1} {\cal L} \ =\ - \ \frac12 \, R \ - \ \frac12 \,\partial^m \phi \,\partial_m \phi \ - \ V(\phi) \ . \label{min_coupled}
\ee
Here we have set $M_P=1$, a convention that we shall abide to in the remainder,
$e = \det\left(e_m^a\right)$ is the determinant of the metric vielbein and $V(\phi)$ is the scalar potential of the inflaton field.
Inflation occurs in regions where the scalar energy density is dominated by a very flat potential energy. A maximally symmetric metric is then of de Sitter form
\be
ds^2_\text{inflation} \ =\  -\ dt^2 \ + \ a(t)^2 d\vec{x}^{\,2} \ , \qquad a(t) \sim a_0 \,e^{Ht}  \,,
\ee
with $a(t)$ the scale-factor and $H$ the Hubble scale during inflation, which is determined by the value of the scalar potential according to
\begin{equation}
V|_\text{inflation} \ \simeq \ 3 \,H^2 \ .
\end{equation}

The inflationary phase is characterized by the so--called slow--roll conditions, which are usually formulated in terms of the two parameters 
\be
\epsilon \ = \ \frac12 \left( \frac{V'}{V} \right)^2  
 \quad , \qquad
 \eta  \ = \  \frac{V''}{V}  \ ,
\ee
where $V'$ and $V''$ denote derivatives of the potential with respect to $\phi$. 
Inflation occurs when the slow-roll parameters satisfy
\be
\epsilon   \ll 1 \quad , \qquad |\eta |  \ll 1 \ ,  
\ee 
and ends when they attain ${\cal O}(1)$ values. 
The exponential growth of the scale-factor (which one typically quantifies via the number $N$ of e-folds, the logarithm of the ratio between the scale factors at the end and at the beginning of inflation) is dominated by the deep inflationary regime, where the slow--roll parameters are very small; towards the end of inflation, a few final e-folds can be over (or under) estimated by the convenient slow--roll treatment.

A viable inflationary model should provide about 60 e-folds, while satisfying key
constraints from the CMB.
In the slow--roll approximation, the number of e-folds can be expressed as
\be
N_\star  \ = \ \int_{t_\star}^{t_\text{end}} H dt \ \simeq \ \int_{\phi_\text{end}}^{\phi_\star} \frac{V}{V'} \,d\phi \,,
\ee
where $\phi_\star$ is the value of the inflaton at the so-called pivot scale. This corresponds to the horizon exit of typical scalar perturbations observed in the CMB, where one places restrictions on $\epsilon$ and $\eta$.

The slow-roll parameters are used to quantify the constraints on model building arising from CMB observations.
Demanding that $N_\star \sim 55-60$,~\footnote{Actually, the minimum number of e-folds depends on the inflation scale $M_I$. An approximate formula is $N\gtrsim \ln{\frac{M_I}{\rm eV}}$.} one can estimate the value of the inflaton at the pivot scale for different models and then compute the corresponding values of the two slow-roll parameters $\epsilon|_\star=\epsilon(\phi_\star)$
and $\eta|_\star$.
These values determine the spectral index $n_s$ for scalar perturbations and the so-called tensor-to-scalar ratio $r$ of primordial perturbations at the pivot scale as
\be
n_s \ = \  1\ - \ 6 \epsilon|_\star \ + \  2 \eta|_\star  \quad , \qquad r\ = \ 16\, \epsilon|_\star \,.
\ee

\begin{figure}[t]
\centering
\includegraphics[width=70mm]{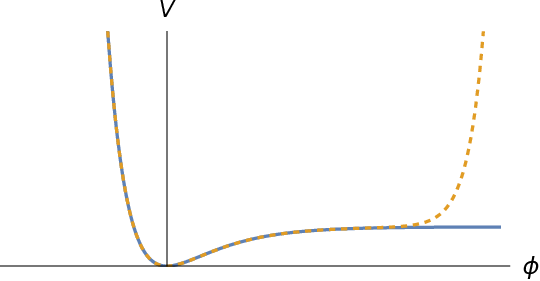}
\caption{\small A standard Starobinsky potential (solid line) and its modification by the inclusion of a tadpole term, as in eq.~\eqref{StaroV+BSB} (dashed line). This forces $\phi$ to descend from the left end and bounce against the steep tadpole wall before inflation starts in the plateau.}
\label{fig:Starobinsky}
\end{figure}

We can now describe the Starobinsky model \cite{Starobinsky:1980te}, which will play an important role in the ensuing discussion.
It is a very instructive setup for inflationary dynamics leading to a small tensor-to-scalar-ratio, which also brings to the forefront some theoretical issues that one is confronted with when attempting to frame inflation within ultra-violet (UV) complete frameworks.
The Starobinsky model can be formulated via a special higher-curvature theory of gravitation, 
\be
e^{-1} {\cal L}\ = \ - \ \frac12  \, R \ +\ \frac{1}{12 \,m^2} \,R^2 \,,
\ee
where $m$ is a dimensionless constant.
A duality transformation can recast it into the more conventional form~\eqref{min_coupled} of gravity minimally coupled to a real scalar field (often referred to as the scalaron), with~\cite{Whitt:1984pd}
\be
\label{StaroV}
V(\phi) \ = \ \frac34\, m^2 \left( 1 \ - \ e^{- \sqrt{\frac23} \phi} \right)^2 \,.
\ee
This potential (the solid line in fig.~\ref{fig:Starobinsky}) grows rapidly for $\phi<0$ and includes an almost flat region for $\phi>0$ where inflation can naturally occur, to then terminate in a reheating phase around $\phi=0$. Typical inflation scales translate into the inequality $m \lesssim 1.3 \times 10^{-5}$.

This setup is clearly very interesting, but the model leaves out a plethora of other curvature--dependent contributions. 
The truncation is difficult to justify from the effective field theory viewpoint. 
In principle this difficulty reflects a more general problem of inflation which has to do with higher--order corrections, and is generically referred to as the ``eta-problem''. 
Specifically, in a theory of the form \eqref{GR+INF}, 
attaining very small values of $\eta$ requires a potential with a wide enough almost flat region for the inflaton. 
Quantum corrections of the Coleman--Weinberg type~\cite{coleman-weinberg} depend on the masses of fields, and the $\eta$ parameter is determined by the typical values of the inflaton mass along the potential. 
Small values for $\eta$ would signal small corrections and large ones would signal, instead, large corrections, which should be avoided to maintain a very flat potential. Keeping the inflaton mass small is challenging, as is usually the case for scalar fields, and this makes the $\eta$ parameter difficult to control.

The situation does not improve by considering the embedding within String Theory. 
As we have stated, in invoking the need for the $R^2$ term, one should not forget that other higher--order terms are left out. 
These include the $R^4$ terms that arise in String Theory, but also other quadratic terms like $R_{mn}\,R^{mn}$, which would introduce ghost--like modes~\footnote{The latter problem can be bypassed by the Gauss-Bonnet combination, which is not topological for $D>4$~\cite{zwiebach} and yet contains only interaction terms around flat space.}.
Neither of these exclusions can be simply justified. Of course in String Theory all higher derivative terms should be interpreted only on-shell since the spectrum is determined at the free-string level and cannot change by the interactions. Off-shell descriptions of the effective action are therefore constrained, and at the four-derivative level should appear only via topological invariants such as the Gauss-Bonnet combination~\cite{Antoniadis:1992sa}, which however does not lead to inflation. Therefore, an $R^2$ term cannot appear as such, but could be generated only approximately via a scalar potential that resembles to \eqref{StaroV}~\cite{Antoniadis:2020txn}. 
Therefore, it seems more promising to leave the geometrical setting aside and simply focus on effective potentials that resemble the Starobinsky shape.

Moreover, the scale of inflation, and thus the Starobinsky plateau, are essentially determined by $m$, and should be low compared to the Planck scale. One would thus need that a relatively large homogeneous patch be present before the onset of Starobinsky inflation~\cite{Goldwirth:1991rj,Ijjas:2013vea}. 
One way to resolve this issue would be modifying the potential {\em before}
the pivot scale so that it becomes steeper and takes the chaotic form at early stages \cite{Linde:1983gd}. 
These and other issues also concern other inflationary models, and point to the lack of a microscopic theory underlying inflation. 

On a different note, the ``brane SUSY breaking'' scenario of~\cite{sugimoto,Antoniadis:1999xk,Angelantonj:1999jh,Angelantonj:1999ms,Dudas:2000nv,Pradisi:2001yv} could help to define a unique fate for the dynamics, if the Starobinsky potential were to combine with a four--dimensional relic of the ``tadpole potential''~\eqref{tadpole}. If the internal volume were somehow stabilized, one could then be confronted with a potential of the form
\be
\label{StaroV+BSB}
V(\phi) \ = \ \frac34 \,m^2 \left( 1 \ - \ e^{- \sqrt{\frac23} \phi} \right)^2 \ + \ T \ e^{\sqrt{6} \left(\phi-\phi_0\right)}\ ,
\ee
for a canonically normalized scalar field~\cite{fss}. Here the inflaton is identified with the string dilaton, while this Starobinsky potential, for which we are not providing a top--down justification, would have a non-perturbative origin. 
Still, if the scalar dynamics is traced back to the original singularity, a striking change of behavior occurs precisely for the orientifold tadpole potential: with this exponential or steeper ones, the scalar can only emerge from the singularity while ``climbing up''~\cite{climbing}! Referring to fig.~\ref{fig:Starobinsky}, in this setup the scalar could only descend the Starobinsky potential during a fast pre--inflationary epoch to then bounce against the steep potential before entering the slow-roll phase followed by a final reheating. The initial-value problem would thus find a definite solution, while a depressed CMB quadrupole would naturally emerge from short--enough inflationary epochs. This post-diction is encouraged by the comparison with data presented in~\cite{climbing_cmb}, but there is also an interesting prediction: the tensor-to-scalar ratio would typically increase by about one order of magnitude within the region where the depression occurs. The next generation of CMB experiments will convey important information in this respect.

\subsection{\sc Inflationary model building in supergravity}

As we have seen, inflationary models can rely on a single scalar field whose potential includes an almost flat region, while other scalar fields are well stabilized and thus inert.
There is a handful of proposals for achieving this in supergravity,
and here we shall focus on those closer to non-linear supersymmetry.
The first models of this type were originally formulated in \cite{Kallosh:2010xz}. 
The bosonic sector of an N=1 supergravity relevant to this end is controlled by the K\"ahler potential $K$ and the superpotential $W$, and has the form
\be
e^{-1} {\cal L} \ = \  - \ \frac12 \,R \ - \ g_{i\overline j}\, \partial A^i \,\partial \overline A^{j} \ - \ V(A^i,\overline A^{j}) \,
\ee
where~\cite{CFGVP}
\be
\label{VSG}
V \ = \ e^K \left( \left|D_i\,W\right|^2  \ - \ 3 \left|W\right|^2 \right) \,.
\ee
Here
\be
D_i\,W\ =\ \partial_i\,W \ +\ K_i\, W \, \qquad  \left|D_i\,W\right|^2 \ = \ g^{i \overline j} \,D_i\,W \,\left(D_j \,W\right)^* \ ,
\ee
while the K\"ahler metric is $g_{i \overline j} = K_{i \overline j}$,
and the bosonic fields are the complex scalars $A^i$. More details on the properties of 4D N=1 supergravity and on our conventions can be found in~\cite{Wess:1992cp}.

The model of interest includes two chiral superfields $T$ and $S$, related to the inflaton and to the so-called stabilizer.
The stabilizer has a dual role in the models of~\cite{Kallosh:2010xz}: it generates the scalar potential for the inflaton, while also giving large masses to the other scalar fields.
As a result, the relevant portion of the $S$ superfield is not its physical scalar $S$, but rather its bosonic auxiliary field $F^S$. This component plays a special role when one resorts to non-linear realizations.

The models of \cite{Kallosh:2010xz} contain a superpotential of the form
\be
W \ = \ S\, f(T) \ ,
\ee
and with a proper choice of K\"ahler potential one can build a variety of inflationary models. 
In particular,~\cite{Kallosh:2010xz} demands a shift symmetry for the inflaton component field in the K\"ahler potential, 
which is only broken by the superpotential. 
We will not use the mildly broken shift symmetry of \cite{Kallosh:2010xz} here, but we shall instead rely on the same type of superpotential and on similar assignments for the superfields.

For the Starobinsky model~\cite{Kallosh:2013lkr,Farakos:2013cqa}
\be
\label{STKW}
K \ = \ - \ 3 \log \left[T \ + \ \overline T \ - \ 2 \left|S\right|^2 \ + \ \frac{\zeta}9\, \left|S\right|^4  \right]  \quad , \quad W \ = \ 6\, m\, S\,\left(T-\frac{1}{2}\right) \ ,
\ee
and the exact supergravity extension of the bosonic $R+R^2$ action~\cite{Cecotti:1987sa} is recovered when $S=0$. 
If $\zeta=0$ the scalar component of $S$ becomes tachyonic around $S=0$ and destabilises inflation, while for sufficiently large values of $\zeta$ three out of the four real scalars can be strongly stabilized. In this fashion, during the inflationary phase
\be
S|_\text{inf.} \ = \ 0 \ , \qquad \text{Im}T|_\text{inf.} \ = \ 0 \ ,
\ee
while the fourth real scalar can play the role of the inflaton, with $\text{Re}T|_\text{inf.} = t \gg 1$,
and the effective Lagrangian during inflation takes the form
\be
e^{-1}  {\cal L}|_\text{inf.} \ = \  - \ \frac{1}{2} \,R
\ - \ \frac{3}{4 t^2} \,\partial^m t \,\partial_m t
\ - \ 6\, m^2 \, \frac{(t-\frac{1}{2})^2}{4 t^2}  \ .
\ee
The redefinition
\be
t \ = \  \frac12  \,e^{\sqrt \frac23 \,\phi} 
\ee
casts the scalar potential precisely in the form of~eq.\eqref{StaroV} and the scalar kinetic term in its canonical form. Keeping the other three scalars as spectators was a crucial ingredient of this procedure, and it is justified if their masses are large enough: in this fashion the CMB spectrum is not affected by their presence.  One clearly needs the condition $m_{\text{spectators}}^2 > H^2$, which is meant to hold in this model. A way to attain this scale separation is presented in~\cite{Kallosh:2010xz}.

If the Starobinsky model is regarded as a model of {\it higher curvature supergravity}, the equivalence holds naturally, and was studied in~\cite{Cecotti:1987sa,Farakos:2013cqa}.
There are different off--shell completions of 4D N=1 supergravity, which lead to different types of inflationary models.
The option that we are about to describe includes two chiral superfields, and is linked to the so-called old-minimal version. In contrast, the higher curvature extension of the new-minimal supergravity would lead to a dual theory including a massive vector multiplet coupled to standard supergravity~\cite{Farakos:2013cqa,Cecotti:1987qe}.

The contributions to the scalar potential should satisfy
\be
D_T W|_\text{inf.} \ = \ 0 \quad , \qquad D_S W|_\text{inf.}\ = \ 6 \,m \,t \ne 0 \quad , \qquad W|_\text{inf.} \ = \ 0 \ .
\ee
These conditions imply that, aside from minor effects due to its kinetic energy, the $T$ superfield does not contribute to supersymmetry breaking, which is sourced by the auxiliary field of the $S$ multiplet:
\be
F^T|_\text{inf.}\ = \ 0 \ , \qquad F^S|_\text{inf.}\ \ne\  0 \ .
\ee
When inflation ends, the inflaton approaches the value $t=0$, and supersymmetry is restored.
Taking these properties into account, in the following we shall see how non-linear supersymmetry plays a natural role in this setting, as was first proposed in \cite{Antoniadis:2014oya}.

The embedding of inflationary models in supergravity continues to suffer from the same issues that have already emerged from our discussion, and most notably the eta-problem and the problem of initial conditions.
One may also wonder how the Starobinsky portion of the potential and superpotential might emerge from String Theory. An example is presented in~\cite{Antoniadis:2020txn} for heterotic four-dimensional strings.
The KKLT scenario~\cite{KKLT} provides an alternative setup for inflation in String Theory. Let us stress that its uplift can be ascribed to a tadpole potential as that originating from brane supersymmetry breaking.

\section{\sc Non-linear supersymmetry}

We can now elaborate on the constrained superfield formalism for non-linear supersymmetry and its applications to supergravity.

\subsection{\sc The goldstino sector}

We have seen that the scalar potential of four--dimensional N=1 supergravity has the specific form of eq.~\eqref{VSG}, so that the vacuum energy cannot be positive unless supersymmetry is spontaneously broken. During the inflationary phase one is inevitably confronted with broken supersymmetry, and in addressing it one should focus on the supersymmetry-breaking sector involving the goldstone fermion or goldstino, which we shall denote by $G$.
This can belong to a chiral superfield 
\be
X \ = \ A^X \ + \ \sqrt 2 \,\theta \, G \ + \ \theta^2 \, F^X \ ,
\ee
and we shall readily assume that $\langle F^X \rangle \ne 0$, while $X$ satisfies the nilpotency constraint \cite{Rocek:1978nb,Casalbuoni:1988xh}
\be
\label{X2}
X^2 \ =  \ 0 \ . \label{X20}
\ee

This constraint is imposed on the superfield, independently of the dynamics, as part of the definition of the goldstino sector, and should be solved in a supersymmetric way. The reader may note the analogies with  non-linear realizations of ordinary symmetries, where similar algebraic constraints emerge at low energies.
Since $X^2$ is a chiral superfield, the condition \eqref{X2} comprises the three conditions
\be
X^2 | \ = \ 0 \ , \qquad D_\alpha(X^2) | \ = \  0 \ , \qquad D^2(X^2) |\  = \ 0 \ .
\ee
The third of these is solved linking the first two component fields of $X$ according to
\be
A^X = \frac{G^2}{2 F^X} \,,
\ee
and one can see that the others are mere consistency checks of this result.

A dynamical justification for using the nilpotency condition \eqref{X2} was presented in \cite{Casalbuoni:1988xh}, where it was linked to the decoupling of the complex scalar $A^X$.
It was then realized that generic systems with broken supersymmetry include this nilpotent superfield, independently of microscopic details \cite{Cribiori:2017ngp}.
We shall not elaborate further of how such a constraint emerges in the IR dynamics, but 
we can present an intuitive argument to this effect.
If supersymmetry is spontaneously broken, there is a fermion in the system, say $\lambda_\alpha$, whose variation 
\be
\delta \lambda_\alpha \ = \  \epsilon_\alpha \ + \ \dots ,
\ee
includes an inhomogeneous contribution,
where $\epsilon_\alpha$ is the  (local or global) supersymmetry parameter and the ellipsis indicate higher--order terms that depend on the specific representation of the goldstino.
Resorting to suitable field redefinitions, the fermion $\lambda$ can always be described by
\be
\label{lambdaG}
\lambda_\alpha \ = \ \frac{G_\alpha}{\sqrt 2 F^X} \,,
\ee
since in any case
\be
\delta G_\alpha \ =\  \sqrt 2\, F^X  \epsilon_\alpha \ + \ \dots \ .
\ee
Systems with spontaneously broken supersymmetry can thus involve the nilpotent superfield $X$, in view of the field redefinition \eqref{lambdaG}. The details of this step were analyzed in \cite{Cribiori:2017ngp}.

Due to the nilpotency constraint on $X$, 
the most general Lagrangian (up to higher order superspace terms) that captures the dynamics of the goldstino is given by 
\be
{\cal L}_\text{SUSY} = \int d^4 \theta X \overline X + \left( \int d^2 \theta f X + c.c. \right) \,,
\ee
and takes the form in components 
\be
{\cal L}_\text{SUSY} \ = \ - \ f^2 \ + \ i \partial_m \overline G \, \overline \sigma^m G \ + \ \frac{1}{4 f^2}\, \overline G^2 \partial^2 G^2
\ - \ \frac{1}{16 f^6}\, G^2 \,\overline G^2\, \partial^2 G^2 \,\partial^2 \overline G^2 \, ,
\ee
after integrating out the auxiliary field $F^X$~\cite{Komargodski:2009rz}. 
This action is equivalent to the original Volkov--Akulov one, as shown in \cite{Kuzenko:2010ef}. 
If the system is coupled to supergravity,
the simplest Lagrangian has the form
\be
{\cal L}_\text{SG} \ = \ -\ 3 \int d^4 \theta \,E \,e^{- \,\frac{X \,\overline X}{3}} \ + \ \left( \int d^2 \Theta \  2{\cal E} \left(f X \,+\, W_0 \right) \ + \ c.c. \right) \ , \label{nlsugra1}
\ee
where  $2 {\cal E}$ is a chiral superspace density while $E$ is a real superspace density. 
These fields live in superspace and the lowest component is, in both cases, $e=\sqrt{-g}$. 
However, they transform as densities, so as to provide local supersymmetric invariant actions when combined with superfields. 
The $\theta$'s are standard Grassmann superspace co-ordinates, while the $\Theta$'s are appropriate Grassmann variables such that chiral superfields have a simpler description when expressed in terms of them, see e.g. \cite{Wess:1992cp}. 
Without loss of generality, $f$ is a real constant and $W_0$ a complex constant, and the K\"ahler potential and the superpotential are the most general ones that can be written in terms of the $X$ superfield, due to its nilpotency. 
The component form of this construction is rather lengthy and was analyzed in a series of articles, including~\cite{Bergshoeff:2015tra}. 
However, after integrating out the auxiliary fields, in the unitary gauge $G_\alpha=0$ the Lagrangian reduces to
\be
\label{DSG0}
\begin{aligned}
e^{-1} {\cal L}_\text{SG} |_{G=0} \ =  & \ - \ \frac12 R
\ + \ \frac{1}{2} \epsilon^{klmn} \left(\overline \psi_k \overline \sigma_l {\cal D}_m \psi_n - \psi_k \sigma_l {\cal D}_m \overline \psi_n\right)
\\
&- \ W_0 \, \overline \psi_a \overline \sigma^{ab} \overline \psi_b
\ - \ \overline{W_0} \, \psi_a  \sigma^{ab}  \psi_b
\ - \ f^2 \ + \ 3 \,|W_0|^2  \, .  \label{nlsugra2}
\end{aligned}
\ee
The complex parameter $W_0$ controls the gravitino mass, while $f$ controls the breaking of supersymmetry, 
whose scale is $\sqrt f$. 
The ``super-BEH'' effect that allows to set $G=0$ was further analyzed in \cite{Ferrara:2016ntj}, working in curved backgrounds, also in the presence of matter multiplets. 
One can describe the same system by imposing constraints on the supergravity chiral superfield ${\cal R}$ itself, for example
\be
\left({\cal R} \ - \ \mu\right)^2\ = \ 0 \ .  \label{nlsugra3}
\ee
The constraint (\ref{nlsugra3}) does not eliminate any physical degree of freedom in the gravity multiplet, but it determines algebraically its complex scalar auxiliary field in terms of the other fields. As a result, one finds the component Lagrangian \eqref{DSG0}, after eliminating auxiliary fields. This option was analyzed in~\cite{Dudas:2015eha,Antoniadis:2015ala}, where it was shown that the minimal supergravity action, supplemented with the constraint~(\ref{nlsugra3}) is equivalent, in a dual formulation, to a standard two--derivative supergravity action with
\begin{equation}
K = - 3 \ \log \ (1 + X + {\bar X}) \quad , \quad
W = W_0 + \mu X \ ,  \label{nlsugra4}
\end{equation}
where $X$ is subject to the nilpotency constraint
$X^2=0$.

\subsection{\sc Constraints on other multiplets}

A system with broken supersymmetry can also include other constrained superfields participating in the low--energy dynamics \cite{Komargodski:2009rz}. A simple organizing principle for eliminating decoupled component fields was presented in~\cite{DallAgata:2016syy}.

Before describing the general prescription, let us discuss a simple example presented in~\cite{Brignole:1997pe}. Given a system with two superfields $X$ and $Y$, we would like to demand that its low--energy dynamics does not involve any scalar fields. This can be attained by imposing the constraint
\be
\label{XY}
X\,Y \ = \ 0  \ ,
\ee
together with the nilpotency condition~\eqref{X2} for $X$, which lies at the heart of supersymmetry breaking.
The condition on $Y$ is then solved by a scrutiny of the superspace constraint \eqref{XY} at the component level,
using the expansion
\be
\label{Y}
Y \ =\  A^Y \ + \ \sqrt 2 \,\theta \,\chi^Y \ + \ \theta^2 \,F^Y  \ .
\ee
One can notice that the $D^2 (XY)| = 0$ constraint is solved by
\be
A^Y \ = \ \frac{G \ \chi^Y}{F^X} \ - \ \frac{G^2}{2 (F^X)^2}\ F^Y \ ,
\ee
while the remaining conditions are just consistency checks.
Therefore, the only effect of the constraint \eqref{XY} is to remove the complex scalar $A^Y$ from the spectrum of the low energy theory,
replacing it with a combination of fermions, in such a way that the supersymmetry algebra is respected.
Therefore, low--energy theories with spontaneously broken supersymmetry and only two fermions can be described by superspace Lagrangians only involving $X$ and $Y$~\cite{Antoniadis:2004uk,DallAgata:2015pdd}.

We can now turn to the general rule presented and analyzed in \cite{DallAgata:2016syy}.
Given a superfield $Q$, one can remove its lowest component from the low energy theory by simply imposing the constraint
\be
\label{XXbQ}
X \,\overline X \,Q \ = \ 0 \ .
\ee
A component field in a higher order term of the $\theta$ expansion can be eliminated by introducing the appropriate number of superspace derivatives. For example, the fermion $\chi^Y_\alpha$ in eq.~\eqref{Y} can be eliminated imposing the superspace constraint
\be
X \,\overline X \,D_\alpha Y \ = \ 0 \ .
\ee
This constraint was also analyzed in~\cite{DallAgata:2015zxp}.
One can eliminate more component fields by simply introducing more constraints on a single superfield;
we shall see examples to this effect when we shall discuss inflationary model building.

In conclusion, one can write manifestly supersymmetric Lagrangians by using the types of constrained superfields that we discussed here to describe the low--energy interactions of theories where supersymmetry is  spontaneously broken.
This setup works both in supersymmetry and in supergravity, and can eliminate component fields that do not participate actively in the dynamics at the relevant low--energy scales.
In supergravity, one can implement these conditions to eliminate the auxiliary fields of the supergravity multiplet~\cite{Cribiori:2016qif}, and even the gravitino \cite{Farakos:2017mwd}.

\subsection{\sc Partial breaking of supersymmetry}

Non-linear realizations find their way into models that describe the partial breaking of  supersymmetry. For example, for a 4D N=2 model with partially broken global supersymmetry, there can be a limit where the broken half becomes non-linearly realized in a low energy description, along the lines of what we have already seen.

Non-linear supersymmetry also emerges as a key property of the description of some important BPS objects that appear in String Theory.
In their presence, the breaking of supersymmetry accompanies the partial breaking of spacetime symmetries. For example, the insertion in the vacuum of
a BPS Dp-brane preserves half of the original supersymmetries of the supergravity background, while breaking the rest.
An object of this type is a generalized probe soliton: it does not truncate the bulk spectrum, and in particular does not affect the number of gravitini.
In its presence, the bulk continues to accommodate complete multiplets, while the brane spectrum only contains multiplets with half as many supersymmetries, and
a consistent coupling demands that the broken supersymmetries be non-linearly realized.
Then, combining the appropriate choices of Dp-branes and orientifolds, one can obtain supersymmetric vacua with fewer supersymmetries (as in the SO(32) Type I) or others where supersymmetry is nonlinearly realized (as in the USp(32) BSB model of~\cite{sugimoto}).

We can now discuss a 4D N=2 example with partial supersymmetry breaking that can serve as a good description/approximation for the systems that we just addressed.
Let us focus on situations where 4D N=1 supersymmetry is preserved, thus working with a complete N=1 multiplet, which we shall take to be a vector multiplet.
This multiplet is described by a real superfield $V$, which admits the $\theta$-expansion
\begin{equation}
\label{Vexp}
V \ = \ \ldots \ - \ \theta \sigma^m \overline \theta v_m
\ + \ i \theta^2 \overline \theta \overline \lambda
\ - \ i  \overline \theta^2  \theta \lambda
\ + \ \frac12 \theta^2 \overline \theta^2 \text{D} \,  ,
\end{equation}
where the dots refer to the leading--order terms in $\theta$, which depend on the gauge choice and vanish in the standard Wess--Zumino gauge~\cite{Wess:1992cp}.
The vector $v_m$ is a gauge field, the fermion $\lambda_\alpha$ is the gaugino and D is the real auxiliary field of the vector multiplet.
The field strength of the abelian gauge vector $v_m$ resides in the chiral superfield
\begin{equation}
W_\alpha \ =\  -\ \frac14 \ \overline D^2 D_\alpha V \,.
\end{equation}
For a detailed discussion of the 4D N=1 vector multiplet, the reader can consult~\cite{Wess:1992cp}.
A non-trivial step performed in \cite{Bagger:1996wp} was to define an N=1 chiral superfield as
\be
X \ = \ \frac{W^2}{4m \ +\  \overline D^2 \overline X } \, ,
\ee
where $m$ is a constant that controls the partial breaking of supersymmetry.
This constraint can be solved iteratively to determine $X$ in terms of $W$ and its superspace derivatives, as
\be
X\left(W,DW, D^2W, \dots \right) \ = \ \frac{W^2}{4m} \ + \ \dots \ .
\ee
We refrain from writing down the full superspace result for $X$ here,
which can be found in \cite{Bagger:1996wp}.
The embedding of this constraint in a manifestly N=2 superspace setup was discussed in \cite{Rocek:1997hi,Ferrara:2014oka},
where it was shown that the full 4D N=2 chiral superfield, which can be cast in the form
\be
\chi \  = \ X \ + \ i\,\sqrt{2}\,\tilde{\theta}^\alpha\,W_\alpha \ - \ \tilde{\theta}^2\left(\frac{1}{4}\,\overline{D}^2 \overline{X} \ + \ m\right) \ ,
\ee
satisfies a quadratic nilpotency condition
\be\chi^2 \ = \ 0 \ , \label{N=2quadratic1}
\ee
similar to what we discussed for the N=1 chiral multiplet in N=1 superspace. This constraint simplifies the effective low--energy description of a system involving a complete N=2 vector-multiplet with partially broken 4D N=2 supersymmetry \cite{Antoniadis:1995vb}. The parameter $m$ corresponds to a non-trivial deformation of N=2 superfields and its presence is necessary for realizing partial supersymmetry breaking~\cite{Antoniadis:1995vb,Antoniadis:2017jsk}.
One can also write (\ref{N=2quadratic1}) in an N=1 language, as a combination of two constraints
\be
X^2 = 0 \quad , \quad X W_{\alpha} = 0
\ . \label{N=2quadratic2}
\ee

This setting affords an interesting generalization, discussed in~\cite{dfs}: the cubic constraint
\be
\chi^3 \ = \ 0 \ . \label{N=2cubic1}
\ee
is also solved if $\chi^2=0$, so that it admits a branch of $2 \to 1$ solutions, but can also describe the full $2 \to 0$ breaking of N=2 supersymmetry. In this case, the N=2 constraint
(\ref{N=2cubic1}) can be rephrased, in an N=1 language, as a collection of three constraints:
\be
X^3 = 0 \quad , \quad X^2 W_{\alpha} = 0 \quad ,
\quad X W^{\alpha} W_{\alpha} = X^2
\left(\frac{1}{4} {\bar D}^2 {\bar X} - m\right)
\ . \label{N=2cubic2}
\ee
The chiral fermion $\Psi_X$ and the gaugino $\lambda$ are in this case the two goldstini of supersymmetry breaking.

Returning to the partial breaking, an N=1 system that contains $X$ will have a manifest N=1 supersymmetry.
However, with an appropriate choice of Lagrangian, the system can have a second,
non-manifest supersymmetry that is non-linearly realized and has the form
\begin{equation}
\delta^* W_\alpha \ = \ 2\, m \, \eta_\alpha
\ +\ \frac12 \,\overline D^2 \overline X \, \eta_\alpha
\ + \ 2i \, \partial_{\alpha\dot\alpha} X \, \overline \eta^{\dot\alpha} \, ,
\quad
\delta^* X \ = \  \eta^\alpha W_\alpha \, ,
\end{equation}
where $\eta_\alpha$ denotes the corresponding global spinor parameter.
A minimal Lagrangian that respects the second non-linear supersymmetry can be built from $X$,
and has the form \cite{Bagger:1996wp}
\be
{\cal L}_{\rm BG} \ = \ \frac{1}{m} \int d^2 \theta \ X \ + \ cc\ .
\ee
It is manifestly invariant under the superspace N=1 supersymmetry, but is also invariant under the second non-linear one. Turning to the component form and integrating out the D auxiliary field, one finds
\be
{\cal L}_{\rm BG} \ = \ \left(1 - \sqrt{ - \det \left(n_{mn} + \frac 1mF_{mn}\right) } \right) \ +  \ \text{fermions} \ .
\ee
This result deserves a few comments.
First of all, the system includes fermions, and if one removes the Maxwell field the interactions of these fermions are described by the Volkov--Akulov model (see e.g. \cite{Kallosh:1997aw,Kallosh:2014wsa}).
Moreover, the bosonic sector takes the Born--Infeld form,
a property that is interesting in its own right,
but even more so since the world-volume vector living on a single Dp-brane has a dynamics dictated by this action.
In addition,
the link to D-brane effective actions is sharpened when one takes into account that this system has actually two supersymmetries:
a linear one, which relates the vector to the fermion,
and a non-linear one, for which the fermion serves as a Goldstone mode.
The relation with new types of Fayet--Iliopoulos terms was further discussed in \cite{Cribiori:2018dlc,Antoniadis:2019xwa}.

Non-linear supersymmetry from the partial breaking of N=2 where, at low energies, scalar multiplets rather than vector multiplets survive, was discussed in \cite{Bagger:1997pi,Rocek:1997hi,Antoniadis:2017jsk,Farakos:2018aml}.

\section{\sc Toward an effective supergravity theory of inflation}

In this section we discuss different classes of models of the inflationary phase that rely on constrained superfields.

\subsection{\sc Models with a nilpotent stabilizer superfield}

As we have seen,  most scalar fields (aside from the inflaton, of course) can become heavy and decouple during the inflationary phase.
In our discussion of constrained superfields, we have also seen that, when a scalar becomes heavy and decouples, one can eliminate it from the spectrum using an appropriate constraint, compatibly with the preservation of supersymmetry, albeit in a non--linear phase.
This option was first implemented in cosmological model building in~\cite{Antoniadis:2014oya}, which addressed the Starobinsky model while imposing the nilpotency condition on the stabilizer superfield. This
was indeed the first instance in which constrained superfields were used in Cosmology.
In this case the K\"ahler potential $K$ and the superpotential $W$ read
\be
K \ = \ - \ 3 \,\log \left[T \,+\, \overline T \,-\, 2\, X \,\overline X  \right]  \ , \qquad W \ =\ 6\, m\, X\,T+fX+W_0 \ ,
\ee
and the $X$ goldstino superfield is subject to the quadratic constraint $X^2\ = \ 0$. 
The actual value of the parameter $f$ is redundant: $f$ can indeed be set to $-3m$ by appropriate rescalings (so that the superpotential reduces to the one of \eqref{STKW} with  a vanishing $W_0$, of course). 
The parameter $m$ is instead physical, and controls the scale of inflation, 
while $W_0$ controls the Lagrangian gravitino mass while it does not enter the scalar potential. 
The only scalars that participate actively in the dynamics are the real and imaginary parts of the lowest component of $T$:
the former is the inflaton, while the latter is an axion that is effectively heavy, during inflation, due to the dynamics. In contrast, the $X$ superfield only contributes Fermi modes.

Comparing this system with eq.~\eqref{STKW}, one can see the term $\zeta X^2 \overline X^2$ is inevitably missing, in view of the nilpotency constraint.  This is not a mere technical accident:
the missing term would generate a mass for the scalar of $X$, which would be so large in the present setting that one could ignore it altogether in the low-energy dynamics. One can
therefore think of the constrained setup as resulting from the {\it formal} 
$\zeta \to \infty$ limit. This would impose the condition $X^2 \,\overline X^2 \equiv 0$, which is trivially satisfied if $X^2=0$.
This comment is meant to illustrate that the superspace constraints can be often deduced focusing on the mass terms that would induce the decoupling of specific components. This was discussed, for example, in \cite{Casalbuoni:1988xh,Farakos:2013ih}.

This model approaches a non-supersymmetric Minkowski vacuum at the end of inflation (unless $W_0=0$, in which case the vacuum is supersymmetric), 
where the description in terms of a nilpotent $X$ breaks down because $F^X$ vanishes. 
In order to rely on the same effective description throughout the whole process, the vacuum at the end of inflation should be suitably modified.
A set of versatile models with this property were discussed in \cite{DallAgata:2014qsj}: they all include an inflationary epoch, while supersymmetry remains broken in the $X$ sector at the exit from inflation, albeit with a vanishing vacuum energy.
The models studied in~\cite{Kallosh:2014hxa,Scalisi:2015qga} improved further on this, by introducing a non-vanishing vacuum energy at the end of inflation that can accommodate the current phase of our Universe.
The idea that KKLT-type de Sitter vacua can be constructed resorting to constrained superfields was first proposed in~\cite{Ferrara:2014kva}.

\subsection{\sc Models with a single-scalar constrained superfield}

The success of inflationary models with nilpotent stabilizer superfields led to further investigations based on different types of constrained superfields.
The first efforts utilized a new chiral superfield subject to a constraint that eliminates all component fields but a single scalar \cite{Komargodski:2009rz}.
Cosmological models of this type were studied, for example, in~\cite{Kahn:2015mla,Ferrara:2015tyn,Carrasco:2015iij}.
In these constructions the $X^2=0$ constraint is accompanied by an additional constraint on the chiral superfield that contains the inflaton,
\be
\label{XAAb}
X \left({\cal A} \ - \ \overline{\cal A}\right) \ = \ 0 \ , \qquad \overline{\cal D}_{\dot \alpha}\, {\cal A} \ = \ 0  \,. \label{orthogonal1}
\ee
This condition eliminates all component fields aside from the real part of the lowest component of ${\cal A}$,
$\text{Re}\,{\cal A}|=a$.
One can verify this by deducing that eq.~\eqref{XAAb} is equivalent to three constraints of the form \eqref{XXbQ}:
\begin{eqnarray}
&  |X|^2 \left({\cal A} \ - \ \overline{\cal A}\right) \ = \ 0  \quad &{\rm : \ eliminates } \quad \text{Im}{\cal A}| \ ; \nonumber \\
& |X|^2 \, {\cal D}_{\alpha} \,{\cal A} \ = \ 0
\quad  &{\rm : \ eliminates \ the \ fermion} \quad {\cal D}_{\alpha} {\cal A} | \ ; \nonumber \\
& |X|^2  \,{\cal D}^2 \,{\cal A} \ =  \ 0 \quad &{\rm : \ eliminates \ the \ auxiliary \ field} \quad    {\cal D}^2 {\cal A}| \ .   \label{orthogonal3}
\end{eqnarray}
In global supersymmetry the component fields thus eliminated become functions of the goldstino and the real scalar $a$, compatibly with the supersymmetry algebra.
In supergravity, the component fields of the supergravity multiplet also enter these functions.

These models leave at first sight much freedom for model building.
Indeed, letting
\be
K \ = \ X \,\overline X \ - \ \frac14 \left({\cal A} \ - \ \overline {\cal A}\right)^2 \ , \qquad
W \ = \ g({\cal A}) \ +\ X \,f({\cal A}) \,
\ee
with $\overline{ f(z)} \,=\, f (\overline z)$ and $\overline{ g(z)} \,= \,g (\overline z)$, one
finds the Lagrangian
\begin{eqnarray}
e^{-1} {\cal L}|_{G=0} &=& \frac12\, R
\ + \ \frac{1}{2}\, \epsilon^{klmn} \left(\overline \psi_k \overline \sigma_l {\cal D}_m \psi_n \ - \ \psi_k \sigma_l {\cal D}_m \psi_n\right)
\\[2mm] \nonumber
&-& \frac12 \,\partial^m a \, \partial_m a
\ -\  g(a) \left( \overline \psi_a \overline \sigma^{ab} \overline \psi_b \ + \ \psi_a \sigma^{ab} \psi_b \right)
\ - \ \left(f^2(a) \ - \ 3 \, g^2(a)\right) \, ,
\end{eqnarray}
whose scalar potential can be adjusted to take any form.
There are problematic cosmological implications at the end of inflation~\cite{Hasegawa:2017hgd}, unless the function $g$ is finely tuned.
The constraint of eq.~\eqref{XAAb} eliminates the auxiliary field, among other component fields, and
it was shown in~\cite{DallAgata:2016syy} that this can possibly require non-unitary UV physics. This was confirmed in \cite{Dudas:2021njv}, where it was shown that models based on the ``orthogonal'' constraint of eq.~(\ref{orthogonal1}) have potentially superluminal propagation for the gravitino, and in \cite{Bonnefoy:2022rcw}, where this feature was related to positivity constraints of certain operators in the low-energy goldstino-inflaton Lagrangian. However, these problems disappear if only the first two constraints in (\ref{orthogonal3}) are retained, abandoning the last one that eliminates the auxiliary field.

\subsection{\sc Models with constrained inflaton superfields}

As we have seen, in addition to the scalar partner of the inflaton and the fermion the constraint of eq.~\eqref{XAAb} also removes the auxiliary field from the spectrum, giving rise to the problems of the effective field theory that were highlighted in~\cite{Hasegawa:2017hgd}.
The extra constraint can be avoided following two different proposals discussed in~\cite{Dalianis:2017okk} and in~\cite{Bonnefoy:2022rcw}, which we can now review.

The setup discussed in~\cite{Dalianis:2017okk} aims at constructing models where the inflaton lives in a chiral superfield that is only subject to the constraint
\be
X \,\overline X \left(\Phi \ - \ \overline \Phi\right) \ = \ 0 \ .
\ee
In this fashion,  only the imaginary part of the lowest component of $\Phi$ is removed, while the other component fields are not affected. 
Splitting the lowest component of the inflaton superfield into its real and imaginary parts, according to
\be
\Phi| \ = \  \phi \ +\  i \,b \ ,
\ee
the constraint can be solved to give
\be
\label{bEXPANSION}
b \,=\, - \, i \,\frac{G \chi^\Phi}{\sqrt 2 \, F^X}
\, + \, i\, \frac{\overline G \,\overline \chi^\Phi}{\sqrt 2 \,\overline F^X}
\, + \, i\, \frac{G^2\,F^\Phi}{4 \left(F^X\right)^2}  \, - \, i\, \frac{\overline G^2\,\overline F^\Phi}{4 \big({\overline F}{}^X\big)^2}  -  \frac{G \sigma^c \overline G}{2 |F^X|^2} \  e_c^m \partial_m \phi \ + \ \ldots ,
\ee
where the ellipsis indicate terms containing at least three powers of the goldstino.
Note that eq.~\eqref{bEXPANSION} is the solution in curved superspace,
but the gravitino and the auxiliary fields of the supergravity sector do not contribute to this order.
The structure of this superfield is more transparent when it is written in the unitary gauge, where it takes the form
\be
\Phi|_{G=0} \ = \ \phi \ + \ 2 \, \Theta \, \chi^\Phi \ + \ \Theta^2 F^\Phi \, .
\ee
For model building purposes one can choose, for example,
\be
K \ =\  X \,\overline X  \ - \ \frac14 \left(\Phi \ - \ \overline \Phi \right)^2 \ , \qquad   W \ = \  f(\Phi) \,  X \ + \ g(\Phi) \,  ,
\ee
and then the potential for the canonically normalized real scalar $\phi$ takes the form 
\begin{eqnarray}
\label{XFlagr22}
V \ = \  |f(\phi)|^2 \ +\ 2 \left| g'(\phi) \right|^2 \ - \ 3 \left|g(\phi)\right|^2   \ . 
\end{eqnarray}  
This scalar potential has now the form expected from standard supergravity, 
and in particular the term $2 \left| g'(\phi) \right|^2$ originates from integrating out the auxiliary field $F^\Phi$. 
The bosonic sector in this setup is \emph{minimal}: it only contains gravity and the inflaton.

The inflationary physics of models of this form was studied in \cite{Dalianis:2017okk}, where it is was shown that they provide a versatile framework for model building, while post-inflationary physics does {\it not} suffer from the pathologies that we have previously mentioned.
In particular, the estimate of the gravitino abundance during preheating is along the lines of the indications of standard supergravity.
The gravitino is not overproduced, once a hierarchy between the inflationary scale and the supersymmetry breaking scale due to the nilpotent $X$ superfield is invoked.
The inflationary scale is fixed by the Hubble scale during inflation, but in some models it is
characterized by the inflaton mass $m_\phi$ in the vacuum,
while supersymmetry breaking is characterized by the gravitino mass.
In this case one requires, following \cite{Nilles:2001ry}, that  $m_\phi|_\text{vacuum} \gg m_{3/2}|_\text{vacuum}$
and $\langle F^\Phi \rangle |_\text{vacuum} = 0$ while $\langle F^X \rangle \ne  0$.
However, during the preheating phase supersymmetry breaking is dominated by the inflaton energy density, and the longitudinal component of the gravitino should be identified with the inflatino $\chi_\alpha^\Phi$, assuming $\langle F^\Phi \rangle |_\text{inflation}>>\langle F^X\rangle$,  rather than with the {\it true vacuum goldstino} $G_\alpha$ \cite{Nilles:2001ry}.

In~\cite{Bonnefoy:2022rcw} the fermion partner of the inflaton was also eliminated from the spectrum, with the combined use of the constraints
\be
\label{Im+Ferm}
X \,\overline X \left(\Phi \ - \ \overline \Phi\right) \ =  \ 0 \ , \qquad  X \,\overline X \,{\cal D}_\alpha \Phi \ = \ 0  \,.
\ee
In this fashion, the inflaton superfield is brought to a minimal form, so that in the unitary gauge
\be
\Phi|_{G=0} \ = \ \phi \ + \ \Theta^2 \,F^\Phi \, ,
\ee
while the full solution of the constraints can be found in \cite{Bonnefoy:2022rcw}.
In models with an inflaton superfield of the form~\eqref{Im+Ferm}, the scalar potential has the standard supergravity form, with a {\it minimal spectrum}: a graviton, a massive gravitino and a real inflaton.
For example, one can choose
\be
K \ = \ -\ \frac14\left(\Phi \ - \ \overline \Phi\right)^2 \ +\  |X|^2 \ , \qquad W \ = \ f(\Phi) \,X \ + \ g (\Phi) \ ,
\ee
and then the scalar potential takes the standard supergravity form $V =  |f(\phi)|^2  +  |g'(\phi)|^2  -  3 | g(\phi) |^2$. 
In particular, resorting to the choice proposed in \cite{DallAgata:2014qsj}, one obtains
\be
f \ = \ \sqrt 3\, g \qquad \to \qquad V \ = \  |g'(\phi)|^2 \ ,
\ee
which leaves indeed much freedom for model building.
The physical properties of models with the constraint \eqref{Im+Ferm} were analyzed in \cite{Bonnefoy:2022rcw}, where it was shown that they {\it do not} suffer from physical inconsistencies.

\section{\sc Concluding remarks}

We have reviewed the main properties of constrained superfields and non-linear supersymmetry, focusing on their applications to inflationary model building within four--dimensional N=1 supergravity. 
This formalism leaves out other fields (and certainly the sgoldstino) which are heavier than the Hubble scale of inflation, 
leading to minimal spectra.  
The setup is very versatile and leaves much freedom for model building, but not all the resulting models afford simple UV completions, and potential inconsistencies can surface already within the low-energy theory. 
The same type of formalism could be applied to systems with moduli stabilization and supersymmetry breaking. 
In general, however, by truncating the spectrum one does not see the possible instabilities generated by other heavy scalars.

Lack of space forced us to leave out a more complete treatment of non-linear extended supersymmetry \cite{Kandelakis:1986bj,Cribiori:2016hdz,Kuzenko:2017zla} 
and the issue of goldstino condensation \cite{DallAgata:2022abm,Kallosh:2022fsc,Farakos:2022jcl} (including the related effect on the gravitino \cite{Alexandre:2014lla}).

Future data from observational Cosmology will shed further light on the properties of the CMB and on their microscopic origin, providing further guidance for this type of supergravity model building.

\section*{\sc Acknowledgements}

We thank Anna Ceresole and Gianguido Dall'Agata for the kind invitation to contribute to this volume, and Paolo Natoli for useful comments. 
I.A. was supported by the Second Century Fund (C2F), Chulalongkorn University.
A.S. was supported by INFN (I.S. GSS-Pi) and by Scuola Normale Superiore.

\end{document}